\documentclass{article}

\usepackage{arxiv}
\usepackage{array}
\usepackage{xcolor}

\usepackage{url,hyperref,lineno,microtype,subcaption, graphicx, amsmath, natbib}

\usepackage[utf8]{inputenc} 
\usepackage[T1]{fontenc}    

\usepackage{booktabs}       
\usepackage{nicefrac}       
\usepackage{microtype}      
\usepackage{lipsum}

\newtheorem{insight}{Insight}

\title{Human Cognition through the Lens of Social Engineering Cyberattacks} 

\author{
    Rosana Monta\~{n}ez\ \\
    Department of Computer Science \\
    University of Texas at San Antonio \\
    San Antonio, TX, USA \\
    \texttt{rosana.montanezrodriguez@my.utsa.edu} \\
    \And
    Edward Golob \\
    Department of Psychology \\
    University of Texas at San Antonio \\
    San Antonio, TX, USA \\
    \texttt{edward.golob@utsa.edu} \\
   \And 
    Shouhuai Xu \\
    Department of Computer Science \\
    University of Texas at San Antonio \\
    San Antonio, TX, USA \\
    \texttt{shouhuai.xu@utsa.edu} \\
}

\begin{document}


\maketitle

\begin{abstract}
Social engineering cyberattacks are a major threat because they often prelude sophisticated and devastating cyberattacks. Social engineering cyberattacks are a kind of psychological attack that exploits weaknesses in human cognitive functions. Adequate defense against social engineering cyberattacks requires a deeper understanding of what aspects of human cognition are exploited by these cyberattacks, why humans are susceptible to these cyberattacks, and how we can minimize or at least mitigate their damage. These questions have received some amount of attention but the state-of-the-art understanding is superficial and scattered in the literature. In this paper, we review human cognition through the lens of social engineering cyberattacks. Then, we propose an extended framework of human cognitive functions to accommodate social engineering cyberattacks. We cast existing studies on various aspects of social engineering cyberattacks into the extended framework, while drawing a number of insights that represent the current understanding and shed light on future research directions. The extended framework might inspire future research endeavor towards a new sub-field that can be called {\em Cybersecurity Cognitive Psychology}, which tailors or adapts principles of Cognitive Psychology to the cybersecurity domain while embracing new notions and concepts that are unique to the cybersecurity domain.
\end{abstract}


\section{Introduction}

Social engineering cyberattacks are a kind of psychological attack that attempts to persuade an individual (i.e., victim) to act as intended by an attacker \citep{mitnick2003wiley,anderson2008wiley}.
These attacks exploit weaknesses in human interactions and behavioral/cultural constructs \cite{indrajit2017ijcsi} and occur in many forms, including phishing, scam, frauds, spams, spear phishing and social media sock puppets \citep{stajano2009acm,linvill2019comphumbeh}.

For example,  in the 2016 U.S. election, attackers used so-called social media sock puppets (also know as Russian Troll) or fictional identities  to influence others' opinions \citep{linvill2019comphumbeh}. 

The effectiveness of current security technologies has made social engineering attacks the gateway to exploiting cyber systems.  Most sophisticated and devastating cyberattacks often start with social engineering cyberattacks, such as spear phishing, where the attacker gains access into an enterprise network \citep{hutchins2011LIIWS}. 
Indeed, \citet{mitnick2003wiley} describe many ways to gain access to secure systems using social engineering cyberattacks. Research in social engineering has mostly focused on understanding and/or detecting the attacks from a technological perspective (e.g., 
detecting phishing emails by analyzing email contents). However, there is no systematic understanding of the psychological components of these attacks, which perhaps explains why these attacks are so prevalent and successful.

This motivates the present study, which aims to systematize human cognition through the lens of social engineering cyberattacks. To the best of our knowledge, this is the first of its kind in filling this void.

\subsection{Our contributions}

In this paper, we make the following contributions. First, we advocate treating social engineering cyberattacks as a particular kind of psychological attack.  This new perspective may be of independent value, even from a psychological point of view, because it lays a foundation for a field that may be called Cybersecurity Cognitive Psychology, which extends and adapts principles of cognitive psychology to satisfy cybersecurity's needs while embracing new notions and concepts that may be unique to the cybersecurity domain. This approach would pave the way for designing effective defenses against social engineering cyberattacks and assuring that they are built based on psychologically valid assumptions. For example, it may be convenient to assume that individuals are willing to participate in defenses against social engineering cyberattacks or that victims are simply reckless. However, these assumptions are questionable because most social engineering cyberattacks are crafted to trigger subconscious, automatic responses from victims while disguising these attacks as legitimate requests. 

Second, as a first step towards the ultimate Cybersecurity Cognitive Psychology, we propose extending the standard framework of human cognition to accommodate social engineering cyberattacks. This framework can accommodate the literature studying various aspects of social engineering cyberattacks. In particular, the framework leads to a quantitative representation for mathematically characterizing {\em persuasion}, which is a core concept in the emerging Cybersecurity Cognitive Psychology and it is key for understanding {\em behavior} in the traditional framework of human cognition. Some of our findings are highlighted as follows: (i) a high cognitive workload, 
a high degree of stress, a low degree of attentional vigilance,
a lack of domain knowledge, and/or a lack of past experience makes one more susceptible to social engineering cyberattacks;
(ii) awareness or gender alone does not necessarily reduce one's susceptibility to social engineering cyberattacks;
(iii) cultural background does affect one's susceptibility to social engineering cyberattacks;
(iv) the more infrequent the social engineering cyberattacks, the higher susceptibility to these attacks; (v)
for training to be effective, it should capitalize on high capacity unconscious processing,
with the goal of creating a warning system that operates in parallel with the user's conscious focus of attention;
(vi) it is currently not clear how personality affects one's susceptibility to social engineering cyberattacks; and 
(vii) more studies, especially quantitative studies, need to be conducted to draw better and consistent results.
In addition to these findings, we propose a range of future research directions, with emphasis on quantifying the effect of model parameters (i.e., victim's short-term cognition factors, long-term cognition factors, long-memory, and attacker effort) on the amount of persuasion experienced by the human target.

\subsection{Related Work}\label{s.relate}

To the best of our knowledge, we are the first to systematically explore the psychological foundation of social engineering cyberattacks. 
As discussed in the main body of the present paper, most prior studies focus on social engineering cyberattack or
cyberdefense techniques. For example, \citet{gupta2016ieee} investigate defenses against phishing attacks; \citet{abass2018jis} discusses social engineering cyberattacks and non-technical defenses against them.

Few prior studies have an aim that is similar to ours. \citet{salahdine2019future} review social engineering cyberattacks and mitigation strategies, but they do not discuss factors such as human cognition. \citet{darwish2012ieee} discuss at a high-level the relationship between human factors, social engineering cyberattacks,  and cyberdefenses, but they neither examine what makes an individual susceptible to social engineering cyberattacks nor do they discuss the effect of a victim's psychological and situational conditions (e.g., culture and short-term factors) on the outcome. \citet{pfleeger2012comp_sec} take a multidisciplinary approach to examine cybersecurity from a Behavioral Science perspective, but they do not offer any systematic framework of looking at human cognition in the context of social engineering cyberattacks. The lack of studies in social engineering cyberattacks might be associated with these studies involving human subjects. In an academic setting, approval for deceptive studies on human subjects requires consent from all entities involved, including ethics board and IT department \citep{vishwanath2011_dssystems}. The nature of the topic might also raise sensitivities among those involved \citep{jagatic2007acm}, which can lengthen the process. This can be discouraging for most researchers.

\subsection{Paper Outline}
Section \ref{sec:cognitive-process} reviews a basic framework for human cognition (without the presence of social engineering cyberattacks). Section \ref{sec:victim-cognition-functions} extends the basic framework to accommodate social engineering cyberattacks and systematizes victim's cognition through the lens of social engineering cyberattacks, with future research directions.

Section \ref{s.concl} concludes the present paper.

\section{Overview of Human Cognition}
\label{sec:cognitive-process}

In this section, we review human cognition functions prior to the presence of social engineering cyberattacks. This framework of human cognition serves as a basis for exploring how victims' cognition functions are exploited to wage social engineering cyberattacks.

\subsection{Human Cognitive Functions}
\label{sec:human-cognition-function-wo-cybersecurity}

The term “cognition” can have radically different meanings in different contexts.  Here we use the term “cognition” in the broadest sense - as a descriptive term for the software counterpart to the brain as hardware.  That is, cognition is the abstract information processing that is being implemented by neurons in the brain \citep{pinker2009ww}. From this perspective, cognition can also include information processing that computes emotions as well as the vast majority of neural information processing that is not reflected in our conscious awareness \citep{baars1997jcs}.

Cognitive psychologists often consider information processing to be the basic function of the brain, in the same way that the liver functions as a complex filter and arteries and veins are essentially pipes.  Correlates of information processing in the brain can be directly observed using various methods to record electrical and chemical activity \citep{kandel2000mcgraw}.  Information processing is evident at multiple spatial, from compartments within individual neurons to tightly organized networks having members in different parts of the brain.  These concrete, physically measurable, neurophysiological activities are analogous to the hardware of a computer.  Indeed, neurons have been profitably studied in terms of functioning as Boolean logic gates, and action potentials, perhaps the most characteristic property of neurons, is convey all-or-none binary states \citep{shepherd2004oxford}. 

\begin{figure}[!htpb]
    \centering
    \includegraphics[width=0.8\textwidth,height=0.9\textheight,keepaspectratio]{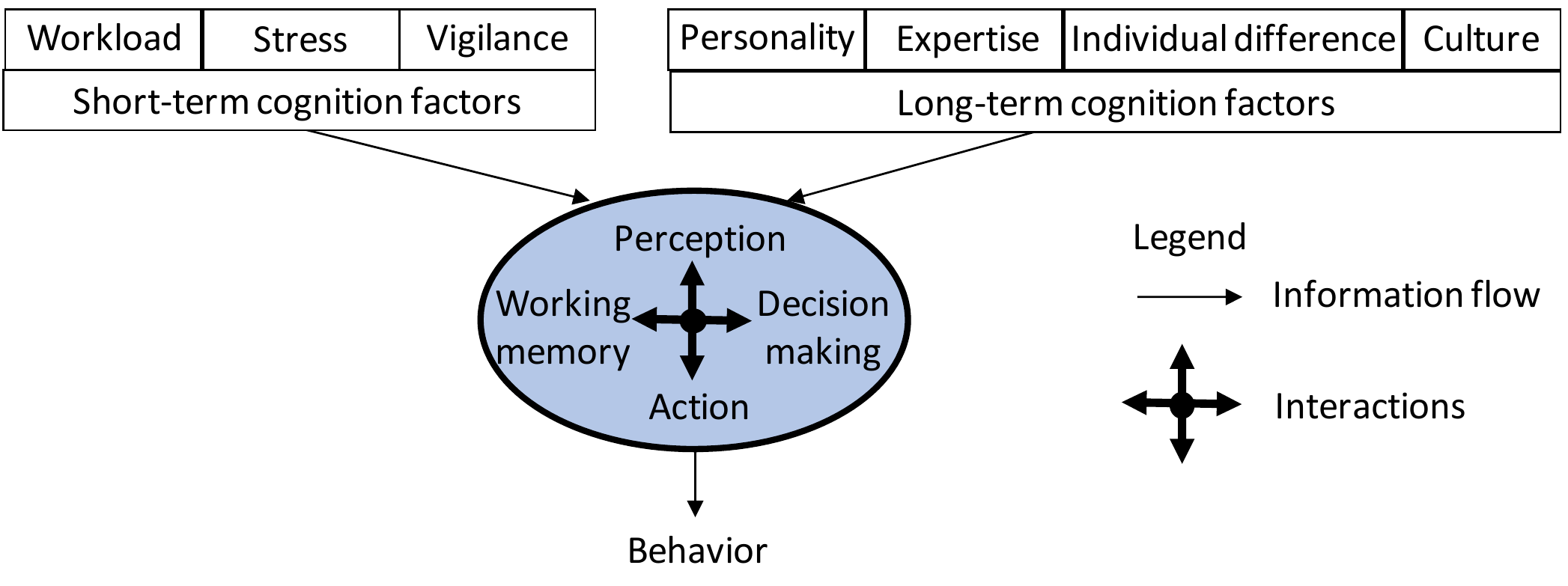}
\caption{A basic, selective schema of human cognition, where the blue background within the oval indicates long-term memory that is accessible by the four components of human cognition functions.}
    \label{fig:cognitve_process}
\end{figure}

Figure \ref{fig:cognitve_process} presents a very basic, and selective, schematic of human cognition functions, which are centered at four information processing components analogous to software components in an information processing system. These four components are called {\em perception}, {\em working memory}, {\em decision making} and {\em action}. These four components are elaborated below as follows. Perception converts information in the word, sampled from the senses, into neural codes that can be used for intelligent behavior and conscious experience \citep{mesulam1998jon}.  Working memory consists of attention and short-term memory, and coordinates information processing by prioritizing certain information for short periods of time, often to accomplish a goal \citep{miyake1999cup}. Decision making further prioritizes information from working memory and other unconscious sources, and is a gateway to behavior \citep{kahneman2011thinking}.  Action is the implementation of computations from decision making, as well as other influences, and also organizes the physical activity of muscles and glands that are measurable as behavior \citep{franklin2011neuron}.  Perception, working memory, decision making, and action are often considered to be roughly sequential, as when trying to hit a baseball, but can mutually influence each other in many ways.  All of these cognitive processes operate on a foundation of accumulated knowledge in memory, which informs these processes, such as when perceiving a familiar face. 

Memory is intrinsic to cognition, because information processing occurs over time and thus requires some information to be retained over time.  The basic processes of perception, working memory, decision making, and action that are engaged “in the moment” use information that is preserved from earlier moments in time.  Memory consists of distinct systems \citep{tulving2000oxford}, in the same way that our domain of “perception” includes the visual, auditory, somatosensory, olfactory, gustatory, and vestibular systems.  One important distinction among systems is whether the information is retained over short periods of time, typically seconds to minutes, or longer periods of time.  In our overview shown in Figure \ref{fig:cognitve_process}, short-term memory is a component of working memory.  Long-term memory contributes to cognition in general, and for this reason, we have situated all of four domains supporting cognition in the moment within long-term memory (indicated by the blue background).  As with the other cognitive domains, memory systems can work in parallel.  For example, the memory of the previous sentence is supported by short-term memory, yet the memory for what each word means resides in long-term memory.

Above, we presented several basic types of information processing that together generate behavior.  We now consider how these basic cognitive processes can be influenced, for better or worse, by a few important factors that are demonstrably relevant to cybersecurity.  The “short-term” factors, reflecting the immediate situation, and “long-term” factors 
are ultimately coded in some form by the brain and exert an influence on the basic cognitive processes that drive behavior. The short-term and long-term factors are elaborated in the next two subsections.  

\subsection{Short-Term Cognitive Factors} \label{s.short}

We focus on three short-term factors: {\em workload}, {\em stress}, and {\em vigilance}.
These factors operate on relatively short timescales (minutes to hours) that have been intensively studied because they impair human performance.  We will consider how these factors may relate to social engineering, and point out the extant literature and promising future directions.

\subsubsection{Workload} 
Human cognition is affected by cognitive workload,

which depends on task demand and the operator in question.

Depending on the details, two tasks can be done at the same time with little or no performance costs (a manageable workload) or be nearly impossible to do well together (a very high workload).  A nice example comes from \citep{shaffer1975qje}, who found that typists could very accurately read and type while they also verbally repeated a spoken message.  Performance, however, plummeted on both tasks if they tried to take dictation (typing a spoken message) while also trying to read a written message out loud.  The differences are thought to reflect the use of phonological (sound-based) and orthographic (visual letter-based) cognitive codes.  In the first example one code is used per task (phonological: listen to speech-talk; orthographic: read-type), while in the second each code is used for both tasks (speech-type; read-talk).  To account for these complexities, psychologists have developed theories that consider different types of cognitive codes 

\citep{navon1979psyrev}, such as auditory or visual sensory input, higher-level verbal or spatial codes, and output codes for driving speech or manual behaviors 

\citep{wickens2008humfac}.  Measures have also been developed to quantify the subjective sense of how much “cognitive work” is being done in a given task. Perhaps the most common instrument to measure subjective workload is the NASA-TLX, which has six dimensions that are clearly explained to the subject, such as “mental demand” or temporal demand (time pressure), and are rated on a scale from low to high.  Lastly, neurophysiological measures are often used to provide objective, convergent measures of workload as well as suggest potential neural mechanisms.  Neurophysiological measures such as transcranial Doppler measures of blood flow velocity in the brain, EEG measures of brain electrical potentials, autonomic nervous system activity such as skin conductance and heart rate and its variability, and functional magnetic resonance imaging (MRI) to quantify changes in blood flow that are secondary to neural activity are commonly used \citep{parasuraman2008oxford}.

\subsubsection{Stress} 

Acute stress may also influence cognition and behaviors that are relevant to cybersecurity.  We distinguish acute from chronic stress, with chronic stress beginning after a duration on the order of months, as their impact on cognition can differ and chronic stress is better classified here as a long-term factor.  The neurobiological and hormonal responses to a stressful event have been well studied, as have their impact on behavior \citep{lupien2009nrn}.  Acute stress can influence attention, a vital component to working memory, in ways that are beneficial as well as detrimental \citep{al2002psyphy}.  Attentional tunneling is one such effect of acute stress where attention is hyper-focused on aspects relevant to the cause of the stress, but is less sensitive to other information.  The term tunneling derives from the use of spatial attention tasks, where arousal due to stress leads to subjects ignoring things that are more distant from the focus of attention \citep{mather2011pps}.  In the realm of cyber security, attention tunneling from an emotion-charged phishing message could lead one to hyper-focusing on the email text but ignore a suspicious address or warnings at the periphery.  Working memory is also vulnerable to acute stress 

\citep{schwabe2013tcs}, particularly by way of interfering with prefrontal cortex function 

\citep{arnsten2009nrn,elzinga2005behneu}.  Decision making can be driven in two fundamentally different ways \citep{evans2008arp}.  The first is by relatively automatic processes that are fast but may not be the optimal choice in some instances (termed “heuristics” and “biases”) 

\citep{tversky1974sci,gigerenzer2008pp}.

The second approach is by using conscious, controlled processing reasoning, which is slower but can be more sensitive to the particulars of a given situation.  Acute stress has a variety of effects on decision making and many subtleties 

\citep{starcke2012nbr}, but in general, can impair rational decision making, and one way is by reducing the likelihood of controlled decision making and increasing the use of automatic processing.

\subsubsection{Vigilance}\label{ss.vigilance}
Vigilance and sustained attention are two closely related, sometimes synonymous, terms for the concept that cognitive performance will systematically change the longer you perform a given task.  Here we will use the term vigilance, which in the laboratory is studied in sessions that typically last ~30-60 min.  In a classic work by 

\citep{mackworth1948qje}, subject watched an analog clock and responded to intermittent jumps in the clock hand. Much work since then has showed that performance in a wide range of tasks declines substantially over these relatively short periods of time (termed the “vigilance decrement)

\citep{parasuraman2008oxford}.  In our view, the potential impact of the vigilance decrement on behavior is an important factor to explore, because the probability of user error may covary with time on task.  For example, the likelihood of downloading malware may increase as users go through their email inbox, particularly if they have limited time.
Lastly, we note that although the situational categories of workload, stress, and vigilance are individually important to examine in the realm of cybersecurity, they are also known to interact with each other.  For example, a high workload and prolonged vigilance are stressful 

\citep{parasuraman2008oxford}.  Another distinction to keep in mind is that many laboratory vigilance tasks are boring and have a low workload.  The extent that the vigilance literature generalizes to other settings such as an office, where workers may have high workloads and stress from complex job demands, is an empirical question worth considering in future cybersecurity studies.

\subsection{Long Term Cognitive Factors}\label{s.long}

In contrast to short-term factors that reflect the current situation and can change rapidly, our second grouping of “long-term factors” covers more stable attributes of a person and their experiences that only gradually change.  We consider factors of personality, expertise, age and gender, and culture.  We include personality as a long-term factor, even though it can be situation-dependent as well (as with short-term factors)

\citep{kenrick1988ap}.  These factors offer some predictability of individual behavior in a given situation.  In the context of cybersecurity, long-term psychological factors can impact how an individual responds to social engineering attacks.
 
\subsubsection{Personality} To Psychologists, “personality” is a technical term that differs somewhat from ordinary usage.  It refers to individual differences in thoughts, feelings, and behaviors that are relatively consistent over time and situations. We say "relatively" because, as noted above, thoughts, feelings, and behaviors are highly dependent on the situation, and lifespan approaches have defined notable changes in personality with age 

\citep{donnellan2009chpp}.  Personality research is dominated by the Big 5 framework of personality domains, which was developed over much of the 20th century in various forms 

\citep{digman1997jps}. The Big 5 framework is based on statistical methods (factor analysis) that identify abstract dimensions that can economically account for much of the variance in personality measures.  The factors are labelled conscientiousness, agreeableness, neuroticism, openness to experience, and extraversion.  For present purposes, the labels of the factors are adequate descriptions of the underlying constructs.  Many studies on the relationship between social engineering and personality focus on openness, conscientiousness, and neuroticism which are thought to have the most impact on susceptibility to social engineering. The factors that comprise the Big 5 framework are:
\begin{enumerate}
\item Openness: the willingness to experience new things.
\item Conscientiousness: favors-norms, exhibiting self-control and self-discipline, competence. 

\item Extraversion: being more friendly, outgoing, interactive with more people.
\item Agreeableness: being cooperative, eager to help others, believe in reciprocity.
\item Neuroticism: tendency to experience negative feelings, guilt, disgust, anger, fear, and sadness.
\end{enumerate}

\subsubsection{Expertise} 

Expertise is typically limited to relatively narrow domain and does not transfer to other areas as much as we tend to believe (termed the “transfer problem”) \citep{kimball2000ohm}.  Limited transfer of expertise can be compounded by cognitive illusions such as the Dunning-Kruger effect.  The Dunning-Kruger effect empirically shows that individuals often overestimate their competence relative to their objective performance

\citep{kruger1999jpsp}.  Similarly, the “illusion of knowledge” shows that people generally know far less about a topic than they believe, as revealed by questioning 

\citep{keil2003tcs}. In the realm of cybersecurity, these and other empirical phenomena underpin user over confidence .  As will be detailed below, narrow expertise about cybersecurity can be beneficial, but computer expertise more generally may not confer security benefits.

\subsubsection{Individual Differences}

There are many kinds of individual differences and we focus on two kinds: age and sex/gender; others would include role in companies and seniority.
In terms of age, there are dramatic changes in cognitive function and behavioral capacities of children as they develop \citep{damon2006jws}. Considering how youths can safely use computers is a major parenting, education, public policy, and law enforcement challenge.  Social engineering attacks can readily take advantage of the cognitive and emotional vulnerabilities of children, and countermeasures are often quite different than with adults (see below).  Cognition changes throughout the adult lifespan at a less frenetic pace vs. in children, but longer term changes are similarly dramatic 

\citep{park2009arp,salthouse2012arp}.

Declines in fluid intelligence, essentially ones ability to “think on your feet”, are particularly dramatic and have wide implications for everyday life 

\citep{horn1967ap}.  

Overall, there are many changes, some declining with age (fluid intelligence) but others not

\citep{schaie2005rhd}.  Another angle is that age is positively associated with the risk for many neurological disorders that can impair cognition, such as stroke and Alzheimer’s disease \citep{hof2001elsevier}.  Age-related neurological disorders are not considered “normal aging”, but the potential vulnerability of many elders due to brain disease has been well-known to criminals for a long time.  As expected, social engineering attacks are a major problem for this vulnerable population.

Psychology has a long history of studying sex differences, defined by biology (i.e., the presence of two X or one X and one Y chromosome) and gender, which is a social, rather than biological, construct.  In terms of basic cognitive functions such as working memory and decision making, which are typically studied in a neutral laboratory context (such as remembering strings of letters, judging categories of pictures, etc.) there are generally little or no differences between sexes and genders.  There are a few well-documented exceptions, such as males having an advantage for mental spatial rotations \citep{voyer1995apa}.  The situation is quite different when examining cognition in the context of social and emotional factors  \citep{cahill2006npg}.  For our purposes, sex and gender is a basic consideration for social engineering attacks, particularly spear phishing which is tailored to an individual.  Our list could include many other types of individual differences that are useful for social engineering attacks, such as socio-economic class, education, personal interests, job position.  We chose to focus on age and sex/gender because they are prominent topics in the cognition literature and important considerations for cyber security challenges such as spear phishing.

\subsubsection{Culture}
In mainstream cognitive psychology, culture is not a prominent variable, as much of the basic literature studies participants in countries that have predominantly western cultures \citep{arnett2008apa}.  Nonetheless, a wide variety of studies have shown that cultural differences are evident in many aspects of cognition, such as basic perception, language and thought, attention, and reasoning \citep{miyamoto2013oxford}. Culture is an important variable to consider for any social engineering attack.  A phishing email, for example, is unlikely to be effective if the message violates norms of the target’s culture.  We also consider the more specific case of organizational culture in the workplace, because it is highly relevant to employee behavior as it applies to cyber security \citep{bullee2017ics}.  As with all of the other short-and long-term variables that we consider, culture is assumed to interact with the other variables, with particularly large interactions with age, gender, and perhaps personality.

\section{Victim Cognition through the Lens of Social Engineering Cyberattacks}
\label{sec:victim-cognition-functions}

Social engineering cyberattacks are a type of psychological attack that exploits human cognition functions to persuade an individual (i.e., victim) to comply with an attacker's request \citep{anderson2008wiley}. These attacks are centered around a social engineering message crafted by an attacker with the intent of persuading a victim to act as desired by the attacker. These attacks often leverage behavioral and cultural constructs to manipulate a victim into making a decision based on satisfaction (gratification), rather than based on the best result (optimization) \citep{indrajit2017ijcsi,kahneman2011thinking}. 

For example, one behavioral construct is that most individuals would trade privacy for convenience, or bargain release of information for a reward \citep{acquisti2005ieee}.  

\begin{figure}[!htpb]
    \centering
    \includegraphics[width=0.9\textwidth,height=0.9\textheight,keepaspectratio]{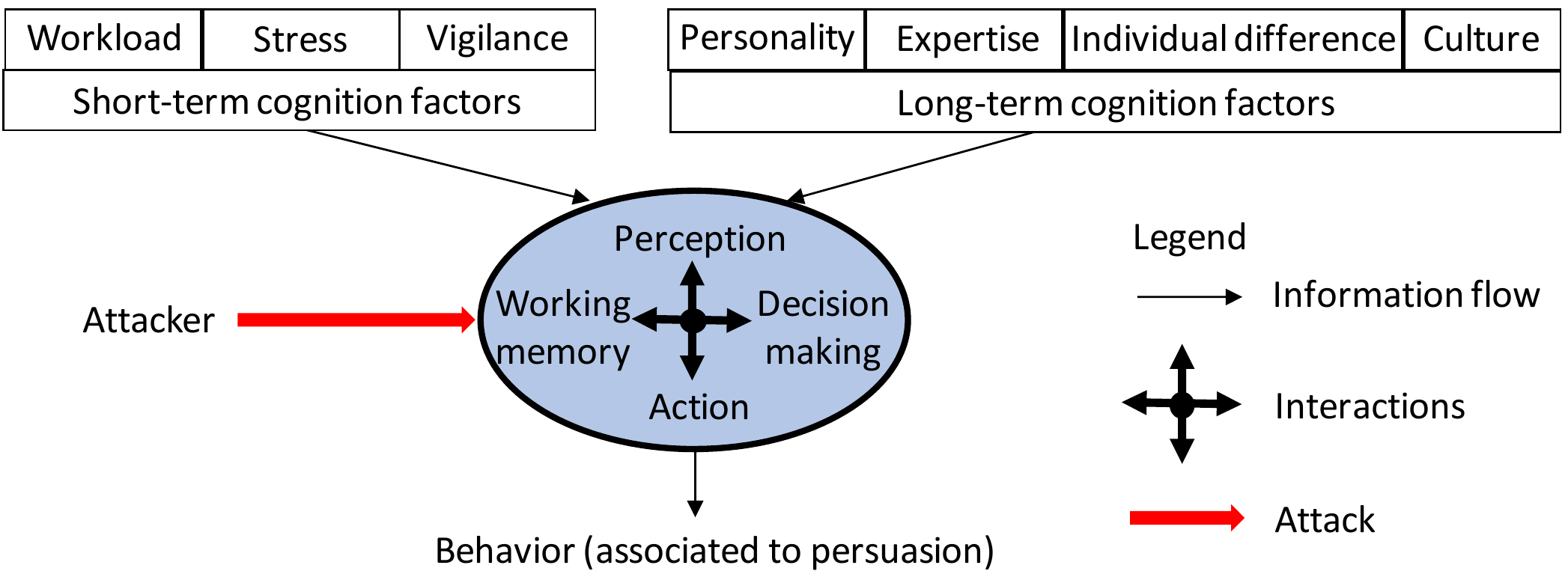}
    \caption{Extending the basic schema of human cognition presented in Figure \ref{fig:cognitve_process} to accommodate social engineering cyberattacks. The extension is to incorporate an attacker that wages a social engineering cyberattack against a victim’s human cognition functions (i.e., the oval). The resulting behavior is associated to persuasion (i.e., an attack succeeds when a victim is persuaded to act as intended by the attacker)..}
    \label{fig:cognitve_process-cybersecurity}
\end{figure}

To establish a systematic understanding of the victim’s cognition through the lens of social engineering cyberattacks, we propose extending the framework presented in Figure \ref{fig:cognitve_process} to accommodate social engineering cyberattacks against human victims’ cognition functions, leading to the framework highlighted in

Figure \ref{fig:cognitve_process-cybersecurity}.  This implies that the resulting behavior of a victim will also depend on the attacker’s effort. In what follows, we will cast the social engineering cyberattacks literature into this framework, by first discussing the literature related to short-term and long-term cognition factors, and then the literature related to cognition functions.

\subsection{Short-term Cognition Factors through the Lens of Social Engineering Cyberattacks}

\subsubsection{Workload}

In computer-mediated communications, cognitive workload can affect an individual's ability to process socially engineered messages.  \citet{pfleeger2012comp_sec} observe that cognitive workload could make individuals overlook elements that are not associated with the primary task. This effect, called inattentional blindness, affects an individual’s ability to notice unexpected events when focusing on the primary task \citep{simons2000tcs}. In most cases, security is a secondary task. For example, when an employee attempts to manage several tasks simultaneously (e.g., reply to hundreds of emails in the email inbox while answering calls and an occasional request from the boss), the employee is more likely to overlook cues in phishing messages that might indicate deception. A study that examined actual phishing behavior by sending employees an innocuous phishing email, found that self-perceived work overload was positively associated with the likelihood of clicking on the phishing link \citep{jalali2020jmir}.
\citet{vishwanath2011_dssystems} investigate the effect of information processing and user’s vulnerability to phishing. Leveraging two phishing attacks that target a university, they survey undergraduate students on their recollection and response to the phishing emails. They find that in the presence of a perceived relevant email, individuals focus more on urgency cues, while overlooking deception cues in the message, such as sender’s email address or email grammar spelling. They also find that individuals that regularly manage large volumes of emails have a high inattentiveness when evaluating emails, making them more vulnerable to phishing attacks. They also find that a high {\em email load} triggers an automatic response, meaning that workload significantly increases a victim's vulnerability to phishing attacks.

Summarizing the preceding discussion, we draw:
\begin{insight}
\label{insight:workload}
Cognitive workload, via mechanisms such as inattentional blindness, can increase vulnerability to social engineering cyberattacks. 
\end{insight}

\subsubsection{Stress}

The particular kind of stress, namely acute stress mentioned above, has only been indirectly investigated in the context of social engineering cyberattacks.
\citet{stajano2009acm} examine how principles of scams apply to systems security. Scams are a form of social engineering cyberattack that 
usually involves a physical interaction between attacker and victim. One scamming technique
is the principle of distraction, by which the attacker can take advantage of a victim that is in a state of mind that prevents them from evaluating deceptive cues. For example, when an unemployed individual pays a job recruiting company for job hunting assistance, the individual does not realize that it is a scam. Catphishing is a social engineering cyberattack by which the attacker creates a fictional online persona to lure a victim into a romantic relationship for financial gains. In this case, an individual who is searching for a romantic partner and is experiencing some personal stress might find a catphishing message appealing and, therefore, unable to detect the deception cues in the catphishing messages. 
In summary, we draw:
\begin{insight}
\label{insight:stress}
Stress may reduce one's ability to detect deception cues in social engineering cyberattack messages but the direct effects of acute stress on cybersecurity social engineering have not been examined.
\end{insight}

\subsubsection{Vigilance}

\citet{purkait2014iscs} conduct a study to examine cognitive and behavioral factors that affect user's capabilities in detecting phishing messages. They study attentional vigilance and short-term memory by surveying 621 participants' ability to identify phishing sites, Internet skills, usage and safe practices, and demographics. The measure of "vigilance" was a brief visual search task in six photographs, which did not evaluate vigilance as we conventionally defined it above.  Individual differences in these visual search scores were significant predictors of performance distinguishing spam from phishing websites, which likely reflects the ability to detect visual cues  on the website that distinguish spam from phish sites.

\begin{insight}
\label{insight:vigilance}
Attentional vigilance, particularly the vigilance decrement, may be an important influence on susceptibility to social engineering attacks, but more research is needed. 
\end{insight}

\subsection{Long-term Cognition Factors through the Lens of Social Engineering Cyberattacks}

\subsubsection{Personality}

Personality has been extensively studied in the context of phishing. 
 
Studies show that Big 5 personality traits are related to individuals' susceptibility to social engineering cyberattacks. \citet{pattinson2012imcs} study how personality traits and impulsiveness affect behavioral responses to phishing messages. They find that individuals that score high on extraversion and openness manage phishing emails better than individuals with other personality types.  
\citet{halevi2013acm} find that high neuroticism increases responses to prize phishing messages and that
individuals with a high openness have low security setting on social media account, increasing their exposures to privacy attacks. 
\citet{halevi2016acm} find that personality traits affect security attitudes and behaviors as follows: high conscientiousness is associated to highly secure behaviors, but does not affect self-efficacy (i.e., one's ability in independently resolving computer security issues); high openness is associated to high self-efficacy; high neuroticism is associated to low self-efficacy; and high emotional stability (inverse of neuroticism) is associated to high self-efficacy.

\citet{cho2016ieee} contradict some of the findings presented in \citet{halevi2013acm}, by finding that high neuroticism decreases trust and increases risk perception, which makes one more likely to misclassify benign emails as phishing ones. They also find that higher agreeableness increases trust  and lowers risk perception (i.e., more likely classifying phishing messages as benign). 
Consciousness is commonly associated with self-control, which diminishes impulsive behavior \citep{cho2016ieee}.   \citet{pattinson2012imcs} find that less impulsive individuals manage phishing messages better. 
\citet{halevi2015ssr} show that 
individuals with high consciousness and lower risk perception are more likely to fall victims to social engineering cyberattack messages. 
\citet{lawson2018iergoa} find that extroversion decreases phishing detection accuracy while high consciousness increases detection accuracy, and that openness is associated with higher accuracy in detecting legitimate messages. \citet{darwish2012ieee} find that individuals high in extraversion and agreeableness pose a higher security risk. 
\citet{mcbride2012rti} find that 
consciousness is associated with low self-efficacy and threat severity.
Workman \citep{workman2008asist} and \citet{lawson2018iergoa}
show that personality traits are related to the degree of persuasion by social engineering cyberattacks. 
Summarizing the preceding discussion, we draw:
\begin{insight}
\label{insight:personality}
Literature results are not conclusive on how personality may influence one's susceptibility to social engineering cyberattacks.
\end{insight}

\subsubsection{Expertise} \label{ss.expertise}

Related to expertise, domain knowledge, awareness, and experience have been studied in the literature on their impact on reducing one's susceptibility to social engineering cyberattacks.

\noindent{\bf Impact of domain knowledge}. An individual's knowledge related to cyberattacks increases their capability to resist social engineering cyberattacks.
For example, the knowledge can be about web browsers, including how to view site information and evaluate certificates.
\citet{kumaraguru2006acm} find (i) non-expert individuals consider fewer security indicators (e.g., meaningful signals) than experts; (ii) non-expert individuals used simple rules to determine the legitimacy of a request, while experts also consider other useful information (e.g., context) that may reveal security concerns with the request; (iii) non-expert individuals make decisions based on their emotions, while experts make their decisions based on reasoning; and (iv) non-expert individuals rely more on (spoofable) visual elements to make decisions because they lack the knowledge that security indicators can be compromised, while experts are more efficient at identifying suspicious elements in a message.
For example, corresponding to (iii), they observe that a non-expert individual might decide to download a software program based on how much they {\em want} it and if the downloading website is recognizable;
whereas an expert might consider how much they {\em need} it and if the downloading website is a reputable source.  
These findings resonate with what is found by \citet{klein1991ieee_smc}, namely that experts make decisions based on pattern recognition, rather than purely analyzing the available options. 
\citet{byrne2016chumbeh} find that risk perception for non-expert individuals is influenced by the benefit that can be obtained for an activity, meaning that actions that an individual considers beneficial are performed more often and are perceived as less risky. 

\begin{insight}
\label{insight:domain-knowledge}
Domain knowledge helps reduce vulnerability to social engineering cyberattacks.
\end{insight}

\noindent{\bf Impact of awareness}. As a rule of thumb, training on non-expert individuals often emphasize on awareness.
In a study of victims in frauds involving phishing and malware incidents, \citet{jansen2016ijcybercrime} find that most participants express that they have knowledge of cybersecurity, but it turns out only a few of them indeed have the claimed knowledge. \citet{downs2006acm} find that awareness of security cues in phishing messages does not translate into secure behaviors because most participants are unable to tie their actions to detrimental consequences.  
On the other hand, it may be intuitive that individuals received formal computer education would be less vulnerable to social engineering cyberattacks.
To the contrary, \citet{ovelgonne2017acm} find that software developers are involved in more cyberattack incidents when compared to others. \citet{purkait2014iscs} find that there is no relationship between one's ability to detect phishing sites and one's education and technical backgrounds, Internet skills and hours spent online. 
\citet{halevi2013acm,junger2017comphumbeh,sheng2010acm} find that knowledge acquired through priming and warning does not affect ones' susceptibility to phishing attacks.

\begin{insight}
\label{insight:awareness}
Awareness and general technical knowledge do not necessarily reduce one's susceptibility to social engineering cyberattacks, perhaps because human cognition functions have not been taken into consideration.
\end{insight}

\noindent{\bf Impact of experience}.
\citet{harrison2016Oinforev} find that knowledge about phishing attacks increases one's attention and elaboration when combined with subjective knowledge and experience, and therefore lowers one's susceptibility to fall victim to social engineering cyberattack messages. \citet{wang2012ieee} find that knowledge about phishing attacks increases one's attention to detect indicators.
\citet{pattinson2012imcs} find that the higher familiarity with computers, the higher capability in coping with phishing messages. \citet{wright2010mis} find (i) a combination of knowledge and training is effective against phishing attacks; (ii) individuals with a lower self-efficacy (i.e., one's ability to manage unexpected events) and web experience are more likely to fall victims to social engineering cyberattacks; and (iii) individuals with high self-efficacy are less likely to comply with information requests presented in phishing attacks.
\citet{halevi2016acm} find that a high self-efficacy correlates a better capability to respond to security incidents.  
\citet{arachchilage2014humanbeh} find that self-efficacy, when combined with knowledge about phishing attacks, can lead to effective strategies for coping with phishing attacks. 
\citet{wright2010mis} find that experiential factors (e.g., self-efficacy, knowledge, and web experience) have a bigger effect on individuals' response to phishing attacks than dispositional factors (e.g., the disposition to trust and risk perception). 
\citet{van2017comphumbeh} find that a higher risk perception of online threats is associated with exposure to the knowledge that is specific to the threat. 
\citet{downs2006acm} find that users can detect social engineering cyberattacks that are similar to the ones they have been exposed to.  \citet{redmiles2018acm} find that the more time an individual spends online, the more skilled they are at identifying spams, and the less likely they will click on the links in the spam messages.  \citet{gavett2017plos} find that education and previous experience with phishing attacks increased suspicion on phishing sites. 
\citet{cain2018isapp} find that past security incidents do not significantly affect secure behaviors. \citet{abbasi2016ieee} find (i) older, educated females and males fell victim to phishing attacks in the past are less likely to fall victim to phishing attacks again; (ii) young females with low phishing awareness and previous experience in suffering from small losses caused by phishing attacks do not necessarily have a lower susceptibility to phishing attacks in the future; and (iii) young males with high self-efficacy and phishing awareness and previous experiences in phishing attacks also do not necessarily have a lower susceptibility to phishing attacks in the future. 

\begin{insight}
\label{insight:self-efficacy}
Self-efficacy, knowledge, and previous encounter of social engineering cyberattacks collectively reduce one's susceptibility to social engineering cyberattacks.
In particular, costly phishing experiences would greatly reduce one's susceptibility to social engineering cyberattacks, while non-costly experiences do not.
\end{insight}

\subsubsection{Individual Differences}

Two kinds of individual differences have been investigated in the context of social engineering cyberattacks: gender and age.

\noindent{\bf Impact of gender}.
Initial studies suggest a relationship between gender and phishing susceptibility. 
\citet{hong2013sage} finds that individual differences (e.g., dispositional trust, personality, and gender) are associated with the ability to detect phishing emails. 
\citet{halevi2015ssr} find that for women, there is a positive correlation between conscientiousness and clicking on links and downloading files associated with phishing attacks. \citet{halevi2013acm} find that women exhibit a strong correlation between neurotic personality traits and susceptibility to phishing attacks, but no correlation to any personality trait is found for men.
\citet{halevi2016acm} reports that women exhibit lower self-efficacy than men. \citet{sheng2010acm} find that women with low technical knowledge are more likely to fall victim to phishing attacks. 
\citet{sheng2010acm} find that women are more likely to fall victim to phishing attacks. 

However, later studies provide a different view. 
\citet{sawyer2018humanfactors} finds that there is no relationship between gender and phishing detection accuracy. Similarly, \citet{purkait2014iscs} find that there is no relationship between gender and the ability to detect phishing sites. \citet{byrne2016chumbeh} finds that there is no relationship between gender and risk perception. \citet{rocha2014imcs} finds that there is no significant correlation between phishing resiliency and gender. \citet{bullee2017ics} finds that gender does not contribute to phishing message responses. 
\citet{abbasi2016ieee} finds (i) women with a high self-efficacy have a low susceptibility to social engineering cyberattacks, and that women without awareness of the social engineering cyberattack threat have a high susceptibility to these attacks; and (ii) men with previous costly experiences with phishing attacks have a low susceptibility to these attacks, while overconfidence increases the susceptibility to these attacks. 
\citet{cain2018isapp} find that although men may have more knowledge about cybersecurity than women, there is no difference in terms of insecure behaviors by gender. 
\citet{redmiles2018acm} show that in the context of social media spam, gender affects message appeal but not susceptibility to social engineering cyberattacks,
and that women are more likely to click on sales-oriented spams while men are more likely to click on media spams that feature pornography and violence. \citet{goel2017ais} find that women open more messages on prize reward and course registration than men.
\citet{rocha2014imcs} find that gender affects the type of phishing message an individual would respond to and that women are less susceptible than men to generic phishing messages.

\begin{insight}
\label{insight:gender}
Gender does not have a big impact on the susceptibility to social engineering cyberattacks.
\end{insight}

\noindent{\bf Impact of age}.
Most studies focus on age groups in young people (18-24) and old (45+) ones. In general, youth is related to inexperience, high emotional volatility \citep{zhou2016websyseng}, less education, less exposure to information on social engineering, and fewer years of experience with the Internet. These factors are often accompanied by a low aversion to risk and therefore can increase the chances of falling victim to social engineering cyberattacks \citep{sheng2010acm}. In an experiment involving 53 undergraduate students in the age group of 18-27, \citet{hong2013sage} find that the students' confidence in their ability to detect phishing messages does not correlate to their detection rate. 
\citet{sheng2010acm} investigate the relationship between demographics and susceptibility to phishing attacks and find that individuals at the age group of 18-25 are more susceptible to phishing attacks than other groups 25+. 
\citet{lin2019acm} report a similar result but for an old group. 
\citet{howe2012ieee} find that age also affects risk perception: individuals in the age groups of 18-24 and 50-64 perceive themselves at lower security risks compared to other groups and therefore are more susceptible to social engineering cyberattacks.   
\citet{purkait2014iscs} find that the detection of phishing messages decreases with the age and frequency of online transactions. 
\citet{bullee2017ics} find that age has no effect on spear-phishing responses and that Years of Service (YoS) is a better indicator of victimization (i.e., a greater YoS means less likely susceptible to social engineering cyberattacks).  \citet{gavett2017plos} examine the effect of age on phishing susceptibility
and show that processing speed and planning executive functions affect phishing suspicion, hinting a relationship between phishing susceptibility and cognitive degradation from ageing.

\begin{insight}
\label{insight:age}
Old people with higher education, higher awareness and higher exposure to social engineering cyberattacks are less susceptible to these attacks.
\end{insight}

\subsubsection{Culture}

Culture affects individuals' online activities \citep{sample2018ahfe}, decision making process and uncertainty assessment \citep{chu1999obhdp}, 
development of biases and risk perception \citep{is_subcult,pfleeger2012comp_sec},
reactions to events \citep{rocha2014imcs,hofstede2010mcgraw}, and
self-efficiency  \citep{halevi2016acm,sheng2010acm}.
\citet{redmiles2018acm} suggest that country/communal norms might affect spam consumption as follows: in countries where spam is more prevalent, users are 59\% less likely to click on spam when compared to countries where spam is less prevalent.  

\citet{halevi2016acm} find that individuals with higher risk perception have higher privacy attitudes, which reduce the susceptibility to social engineering cyberattacks. 
\citet{alhamar2010ieee} perform experimental spear-phishing attacks against two groups from Qatar, where one group consists of 129 employees of a company (dubbed {\em employees}) and the other consists of 30 personal acquaintance (dubbed {\em friends}); they find that 
44\% of the individuals in the {\em employees} group are successfully phished while 57\% of the {\em friends} groups are successfully phished. 
\citet{tembe2014acm} 
report that participants from India exhibit a higher susceptibility to phishing attacks when compared with participants from the USA and China. 
\citep{bullee2017ics} report that
participants from China and India might not be aware of the harms and consequences of phishing attacks, while participants from the USA exhibit more awareness of privacy and online security features (i.e., SSL  padlocks) and are more active in safeguarding their personal information. 
\citet{halevi2016acm} 
find that although culture is a significant predictor of privacy attitude, it does not predict security behavior and self-efficacy.

\citet{bohm2011mt} finds that culturally sound messages do not raise suspicion.
\citep{farhat2017linkedin,hofstede2010mcgraw} show that scams with culture-specific shame appeal are more likely to be effective in a certain 
culture. 
\citet{bullee2017ics} 
find that participants from countries with a higher Power Distance Index (PDI), which means that individuals are more likely to comply with hierarchy, are more vulnerable to phishing than those individuals from countries with a lower PDI. 
\citet{Sharevski2019arXirv} show how to leverage cultural factors to tailor message appeal. 

\begin{insight}
\label{insight:culture}
Culture affects privacy and trust attitudes, which indirectly affect one's susceptibility to social engineering cyberattacks.
\end{insight}

\subsection{Victim Cognition Functions through the Lens of Social Engineering Cyberattacks} \label{s.fund}

\subsubsection{Long-term Memory}

As reviewed in Section \ref{sec:human-cognition-function-wo-cybersecurity},
long-term memory is a very broad field, of which the following aspects have been studied through the lens of social engineering cyberattacks.
The first aspect is the {\em frequency of attacks}.
The environment, in which a victim operates, provides a context that may be exploited by an attacker. For example, the attacker may leverage an ongoing societal incident or personal information to craft messages to make the victim trust these messages. 
The attacker attempts to build trust with a victim while noting that a suspicion thwarts the attack \citep{vishwanath2018commresearch}. 
Both trust and suspicion are affected by the environment, such as the frequency of the social engineering events exploited by the social engineered messages. For example, in a situation where social engineering cyberattacks are expected, the attacker is at a disadvantage \citep{redmiles2018acm}; in a situation where social engineering cyberattacks are infrequent, the attacker has the advantage. \citet{sawyer2018humanfactors} investigate how infrequent occurrence of phishing events (i.e., the prevalence of phishing) affects individuals' abilities to detect cyberattacks delivered over emails. In their experiment, they ask three groups to identify malicious and legitimate email messages. The three groups respectively contain 
20\%, 5\%, and 1\% malicious emails. They find that the accuracy of the detection of malicious emails is lower for the group dealing with emails that contain 1\% malicious ones. 
Similarly, Kaivanto \citep{kaivanto2014riskanalysis} show that a lower probability of phishing occurrence increases victim's susceptibility to phishing cyberattacks. 

\begin{insight}
\label{insight:attack-frequency}
The success of social engineering cyberattacks is inversely related to their prevalence.
\end{insight}

Insight \ref{insight:attack-frequency} causes a dilemma: when automated defenses are effective at detecting and filtering most social engineering cyberattacks, the remaining attacks that make it through to users are more likely to succeed.
One approach to dealing with this dilemma is to resort to principles in Cognitive Psychology. It is known that most of the brain’s information processing is sealed-off from conscious awareness \citep{nisbett1977apa}, some permanently while other information could be consciously appreciated, but may not be conscious at a given moment.  Our visual system, for example, computes 3-D depth from 2-D retinal inputs \citep{devalois1990oxford}.  We do not consciously experience the calculations needed to transform the 2-D input into a 3-D percept.  Instead, we are aware of the product (i.e., seeing a 3-D world) but not the process that led to the product.  The influences of subconscious processing are well-known to impact behavior \citep{kahneman2011thinking, nosek2011tcs}. This fact leads to the following insight:
\begin{insight}
\label{insight:training-to-rescue}
Training methods that ask people to consciously think about social engineering cyberattacks are unlikely to be very successful unless the learning reaches the point where it is a habit that, largely unconsciously, guides safer computer use behavior.
\end{insight}

Insight \ref{insight:training-to-rescue} would avoid the dilemma mentioned above
because when the training/learning effort reaches the point that users can deal with social engineering cyberattacks 
subconsciously, users can effectively defend these attacks. This coincides with findings of 
\cite{halevi2015ssr,rocha2014imcs,halevi2016acm,howe2012ieee,sheng2010acm}, namely that users with higher risk perception can reduce the chance they fall victim to social engineering cyberattacks.

\subsubsection{Victim Cognition Functions: A Preliminary Mathematical Representation}

The framework described in Figure \ref{fig:cognitve_process-cybersecurity} formulates a way of thinking in modeling how the behavior of a victim is influenced by the victim's short-term cognition factors (or {\tt short\_term factors}), long-term cognition factors (or {\tt long\_term factors}) and long-memory (or {\tt long\_memory}) as well as the attacker's effort (or {\tt attacker\_effort}). This formulation is applicable to phishing, spear phishing, whaling, water holing, scams, angler phishing and other kinds of social engineering attacks, where the resulting behavior is whether a victim is persuaded by the attacker to act as intended.  For example, spear-phishing is a special case of the model because the attacker often makes a big effort at enhancing {\em message appeal} by exploiting personalization; scam is another special case of the model because the attacker often makes a big effort at enhancing {\em message appeal} by exploiting situational setting and possibly stress. In principle, the behavior ({\tt behavior}) of social engineering cyberattacks can be described as some mathematical function $f$ (mathematically speaking, more likely it will be a family of functions):
\begin{equation}
{\tt behavior}=f({\tt short\_term\_factors}, {\tt long\_term\_factors}, {\tt long\_memory}, {\tt attacker\_effort}).\label{eq:quantitative-model}
\end{equation}
Note that $f$ mathematically abstracts and represents the interactions between the four cognitive domains operating in long-memory (i.e., perception, working memory, decision making, and action), while also taking short-term and long-term factors into account. Moreover, $f$ accommodates attacker's effort as input.
It is an outstanding research problem to identify the appropriate abstractions for representing these model parameters and what kinds of mathematical functions $f$ are appropriate to what kinds of social engineering attacks.
These questions need to be answered using experimental studies. Note that the framework can be expanded to include measures of brain activity, either direct measures such as electroencephalography, or indirectly using peripheral measures such as eye tracking and autonomic nervous system activity \citep{valecha2020isn}.

Specific to the cybersecurity domain, we propose considering persuasion-related behavior, as shown in Figure \ref{fig:cognitve_process-cybersecurity} and the corresponding mathematical representation of Eq. \eqref{eq:quantitative-model}, meaning that the outcome in Eq. \eqref{eq:quantitative-model} can be replaced by the probability that a user is persuaded by the attacker to act as the attacker intended. Intuitively, persuasion is the act of causing someone to change their attitudes, beliefs, or values based on reasoning or argument.  
\citet{wright2014isr} defines Cialdini's Principles of Persuasion, which have been extensively used to study the response to social engineering messages
 (but not social engineering cyberattacks).
Table \ref{table:Cialdini Principles} presents a brief summary of Cialdini's Principles of Persuasion.

{
\begin{table}[!htbp]
\centering
\begin{tabular}{l|p{.83\textwidth}}
 
 \hline
 \textbf{Principle} & \textbf{Description} \\ 
 \hline
 \hline
Liking   & The act of saying yes to something you know and like;
for example, a social engineer presenting himself as helpful and empathetic towards the victim in a password reset process.\\ 
  
\hline
 Reciprocity & Repaying an earlier action in kind; for example, conveying to a victim that they have detected suspicious activities in the victim's credit card account while encouraging the victim to reset the password with their assistance. \\ 
 \hline
 Social Proof & The use of endorsement; for example, stating that due to recent suspicious activities, new security requirements are issued and must be complied by all account holders.\\ 
 \hline
 Consistency & Leveraging the desire of individuals to be consistent with their words, belief, and actions; for example, reminding users that they have to comply with a password reset policy as they have previously done.\\ 
 \hline
 Authority & Responding to others with more experience, knowledge, or power;for example, an email signed by a Senior Vice President of a bank requesting customers to reset their account passwords.\\ 
 \hline
 Scarcity & Something being valuable when it is perceived to be rare or available for a limited time; for example, giving a user 24-hours notice before they deactivate the user's account.\\ 
 \hline
 Unity & Shared identity between the influencer and the influenced\\ 
 \hline
\end{tabular}
\caption{Summary of Cialdini Principles of Persuasion. The principle of Unity was introduced in \citet{cialdini2016ss} but has not been studied in social engineering research; it is presented here for the purpose of completeness.}\label{table:Cialdini Principles}
\end{table}
}

Intuitively, the mathematical function $f$ in Eq. \eqref{eq:quantitative-model} should accommodate or reflect Cialdini's Principles of Persuasion. Although the state-of-the-art does not allow us to draw insights into how these Principles would quantitatively affect the form of $f$, we can still draw some insights from existing studies, as discussed below.

\citet{van2019arXiv} study the relation between phishing attack success and Cialdini Principles of Persuasion using enterprise emails from a financial institution. They find that phishing emails that received the most responses ('clicks') are those who use {\em consistency} and {\em scarcity} principles mentioned above. They also find that emails with more cognitive elements (e.g., proper grammar, personalization, and persuasion elements) receive most responses. 

In a related study, \citet{lin2019acm} find that younger individuals are more susceptible to phishing messages that use the {\em scarcity} and {\em authority} principles, while older individuals are more susceptible to phishing emails that use the {\em reciprocity} and {\em liking} principles. 

\citet{rajin2018frontiers} find that the most successful phishing message strategies are notifications messages, authoritative messages, friend request messages, shared interest messages, and assistance with a failure. These strategies map to Cialdini's Principles of {\em liking}, {\em authority}, and {\em unity}. 

\citet{lawson2018iergoa} find that socially engineered messages

that use {\em authority} and {\em scarcity} principles are considered more suspicious than those that use the {\em liking} principle.

There have been proposals to augment Cialdini's Principles to better represent the psychological vulnerabilities that have been exploited by social engineering cyberattacks. Ferreira and colleagues \citep{ferreira2015ieee_st,ferreira2015humaispt} present five Principles of Persuasion 

by combining (i) Cialdini Principles of Persuasion; (ii) Stajano’s study on scams and how distraction, social compliance, herd, dishonesty, kindness, need and greed, and time affect the persuasive power of scam messages \citep{stajano2009acm}; and (iii) Gragg’s psychological triggers on how strong affect or emotion, overloading, reciprocation, deceptive relationships, diffusion of responsibility and moral duty, authority, and integrity and consistency can influence an individual's response to social engineered messages \citep{gragg2003sans}. Table \ref{table:augmented-principles} presents a summary of these newly proposed five principles.

\begin{table}[!htbp]
\centering
\begin{tabular}{p{.25\textwidth}|p{.7\textwidth}}
 \hline
 \textbf{Principle} & \textbf{Description} \\
 \hline
 \hline
 Authority & Obeying pretense of authority or performing a favor for  an authority.\\
 \hline
 Social Proof & Mimicking behavior of the majority of people.\\
 \hline
 Liking, Similarity, and Deception (LSD) &  Obeying someone a victim knows/likes, or someone similar to the victim, or someone a victim finds attractive.\\
 \hline
 Commitment, Reciprocity, and Consistency (CRC) & Making a victim act as committed, assuring consistency between saying and doing, or returning a favor.\\
 \hline
 Distraction & Focusing on what a victim can gain, need, or lose/miss out. \\ 
 \hline
 \end{tabular}
\caption{Summary of the five newly proposed principles of persuasion that would better represent the psychological vulnerabilities exploited by social engineering cyberattacks \citep{ferreira2015humaispt}.} \label{table:augmented-principles}
\end{table}

Guided by the newly proposed principles, \citet{ferreira2015ieee_st}

conduct experiments, using phishing emails, to show that {\em distraction} (e.g., fear of missing out, scarcity, strong affection, overloading, and time) is the most prevalent phishing tactic,
followed by {\em authority}
and LSD.

Summarizing the preceding discussion, we draw:
\begin{insight}
\label{insight:persuasion-to-f}
The representation of mathematical function $f$ in Eq. \eqref{eq:quantitative-model} should adequately reflect the Principles of Persuasion. 
\end{insight}

Since quantitatively describing the mathematical function $f$, as demanded in Insight \ref{insight:persuasion-to-f}, is beyond the scope of the state-of-the-art, in what follows we explore some qualitative properties of the these mathematical functions, showing how an increase (decrease) in a model parameter would affect the outcome {\em behavior} (more precisely, {\em persuasion}) of a victim. These qualitative observations also need to be quantitatively verified by future experimental studies.

\subsubsection{Impact of Attacker Effort on Victim Behavior}
As shown in Eq. \eqref{eq:quantitative-model}, the attacker can affect victim's behavior through the {\em attack effort} variable, which can be reflected by attacker's {\em message quality} and {\em message appeal} with respect to the victim in question.   
In terms of the impact of message quality,
\citet{downs2006acm} find that most individuals rely on superficial elements when determining if a message is legitimate, without knowing that most of those elements can be spoofed.  
\cite{jansen2016ijcybercrime} find that in online banking frauds involving phishing or malware,  most victims report that fraudulent stories in phishing emails or phone calls appear to be trustworthy.  
Message quality appears increase social engineering cyberattack success. \citet{wang2012ieee} find that {\em visceral triggers} and {\em deception indicators} affect phishing responses. Visceral triggers increases phishing responses, whereas deception indicators have the opposite effect, reducing phishing response.
Similarly,  \citet{vishwanath2011_dssystems} find that individuals use superficial message cues to determine their response to phishing messages. The study reports that urgency cues make it less likely for an individual to detect deception cues.  One common method of trustworthiness is through the use of visual deception, which is effective because most individuals associate professional appearance and logos with a website or message been legitimate. Visual deception involves the use of high-quality superficial attributes (e.g., legitimate logos, professional design, and name spoofing). \citet{hirsh2012psysci} find that phishing messages that use visual deception have a higher victim response rate.  \citet{jakobsson2007privsec} observes that for most individuals, the decision to trust a website or not is based on-site content and not on the status of a site security indicators (e.g., the use security certificates, HTTPS). He also notes that most users could not detect subtle changes in URLs (e.g., a malicious {\tt www.IUCU.com} versus a benign {\tt www.IUCU.org}). Dhamija \citep{dhamija2006acm} conducts a study on malicious website identification. Using malicious and legitimate websites with professional appearance, with the difference that malicious sites display security indicators (e.g., missing padlock icon on the browser, warning on site's digital certificate), they find that 23\% of their participants fail to identify fraudulent sites for 40\% of the time. This group of participants are asked to assess a website's legitimacy based on its appearance (e.g., website design and logos); 90.9\% of participants fail to identify a high-quality malicious website that uses visual deception (i.e., URL spoofing replacing letters in a legitimate URL, for example, "W" for "vv").

On the other hand, message appeal is associated with the benefit an individual derives from complying with a request. \citet{halevi2013acm,halevi2015ssr} find that many individuals that fall to social engineering cyberattacks ignore the risk of their actions because they focus on the potential benefit that the phishing email offers. Message appeal has the most weight on social engineering susceptibility.  An example of this is the Nigerian scam, also known as "419" scam. Herley \citep{herley2012weis} notes that although the scam is well-known and information on the scam is readily available online, individuals still fall victim to it because the message is designed to appeal the most gullible. Two techniques that are commonly used to increase message appeal are \emph{contextualization}, also known as pretexting, and \emph{personalization}. 
\begin{itemize}
\item \emph{Contextualization} is a variation of message framing where the sender provides details or discusses topics relevant to the group to vouch for the victim's membership in the group.  \citet{luo2013cs} conducts a study on phishing victimization with contextualization in the message. In the experiment, they use work benefits and compensation as a pretext in an email to university staff. They find that individuals interpret high-quality content and argument messages (e.g., well written, persuasive messages) as originating from a credible sender. Basing the message argument on a topic that is common within a community (i.e., contextualization) gives the message the appearance of originating from a known person within a group. Using this technique, they are able to achieve 15.24\% victimization in 22 minutes by combining pretexting and message quality.  Similarly, \citet{goel2017ais} examine the effect of messages contextualization (i.e., pretexting) on phishing message opening and compliance rates. They find that highly contextualized messages that target issues relevant to the victim are more successful.
They also find that messages with higher perceived loss have higher success rates than those with high perceived gains. \citet{rajin2018frontiers} also find that phishing messages on work-related or social communication topics (e.g., friend requests) have a higher success rate than messages requesting a password change or offering a deal. 

\item \emph{Personalization} is another framing technique in which a message is tailored to the preference of the victim \citep{hirsh2012psysci}, such as friendship appeals, expressing similar interests. In a phishing experiment, \citep{jagatic2007acm} find that adding personal data found in social networks to phishing emails increased the response rate from 16\% to 72\%. \citet{rocha2014imcs} find that targeted, personalized phishing messages are more effective than generic messages. They find that phishing emails that receive the most responses are those perceived to come from a known source. \citet{bullee2017ics} also find that emails using personalized greeting line were responded 1.7 times more likely when compared with emails with generic greeting lines.  
\end{itemize}

Summarizing the preceding discuss, we draw:
\begin{insight}
\label{insight:message-quality-and-appeal}
Message quality and message appeal, which reflect attacker effort (e.g., using contextualization and personalization), have a significant impact on the attacker's success.
\end{insight}

\subsubsection{Countermeasures against Social Engineering Cyberattacks}

There have been some studies on defending against social engineering cyberattacks.
First, it is intuitive that effective training should heighten a victim's sense of threat because individuals are more cautious and sensitive to detecting elements that might indicate deception. Along this line, \citet{wright2010mis} find that suspicion of humanity was a dispositional factor that increases the detection of deception in phishing messages, more so than risk beliefs and trust. 
\citet{pattinson2012imcs} also find that when individuals participating in experiments 
are aware that a phishing attack is involved, they perform better on detecting phishing emails. Second, \citet{tembe2014acm} find that individuals from the U.S. having higher suspicion and caution attitudes on online communications, when compared to individuals from China and India. 
Third, \citet{vishwanath2018commresearch} find that habitual patterns of email habits and deficient self-regulation reduce viewers' suspicion. Moreover, detecting deception cues decrease social engineering susceptibility.  \citet{kirmani2007marketing} find that detecting persuasion cues in a message activates suspicion and generates a negative response to the message.
Fourth, \citet{canfield2016humanfactors} find that individuals' inability to detect phishing messages increases their susceptibility, regardless of their cautionary behavior.  
For suspicions to be effective in reducing social engineering susceptibility, the risk must out weight the benefit of complying with the message (e.g., message appeal) without affecting decision performance (i.e., accuracy, precision and negative prediction)\citep{cho2016ieee}. \citet{goel2017ais} find that suspicion alone does not prevent phishing victimization because they report that individuals that are suspicious about email messages can still fall victim to social engineering cyberattacks. 
Summarizing the preceding discussion, we draw:
\begin{insight}
\label{insight:defense}
Individual's capabilities in detection social engineering cyberattacks are affected by their awareness of the threats, their cultural backgrounds, and their individual differences in trust/suspicion. 
\end{insight}

Insight \ref{insight:defense} highlights that there is no silver-bullet solution to countering social engineering cyberattacks. On the contrary, effective defense must take into consideration the differences between individuals because they are susceptible to social engineering cyberattacks at different degrees. 

For more effective defenses against social engineering cyberattacks, the following aspects need to be systematically investigated in the future.
\begin{itemize}
\item Achieving effective human-machine interactions in defending against social engineering cyberattacks.
Effective defense would require to (i) detecting message elements that attempt to increase recipients' trust and (ii) increasing recipients' suspicion on messages.  
In either approach, we would need human-machine teaming in detecting social-engineering cyberattacks, highlighting the importance of effective defenses against social engineering cyberattacks.

\item Improving users' immunity to social engineering cyberattacks.
To improve users' immunity to social engineering cyberattacks, we first need to investigate how to quantify their immunity.
In order to improve users' immunity to social engineering cyberattacks, we need to enhance users' {\em protection motivation} and {\em capabilities in detecting deceptive cues}. One approach is to enhance the user-interface design to highlight the security alerts/indicators in email systems and web browsers because their current designs are not effective \citep{schechter2007ieeesp,downs2006acm,abbasi2016ieee,kumaraguru2006acm}. Along this direction,
the user-interface must highlight security alerts/indicators with specific and quantified severity of threats to the user.
The current user-interface design appears to mainly focus on usability and user experience while assuming the presence of (social engineering) cyberattacks as a default. 
This design premise needs to be changed to treating the presence of (social engineering) cyberattacks as the default.
In order to enhance users' capabilities in detecting deceptive cues, 
we need to design new techniques to enhance our capabilities in automatically detecting or assisting users to detect social engineering cyberattacks. Automatic detection has been pursued by previous studies, which however only focus on examining certain message elements that are known to be associated with previous social engineering cyberattacks (i.e., signature-based detectors); these signatures can be easily avoided by attackers. 
In order to possibly detect new social engineering cyberattacks, future detectors should incorporate cognitive psychology elements to detect social engineering cyberattacks, such as the quantification of messages' persuasiveness and deceptiveness.  
In order to design automated techniques to assist users in detecting social engineering cyberattacks, human-machine interaction is an important issue.

\item Achieving human-centric systems design with quantifiable cybersecurity gain. Modern systems design, including security systems design, often focuses on optimizing performance without considering how humans would introduce vulnerabilities while interacting with the system (i.e., assuming away that humans are often the weakest link). One approach to addressing this problem is to change the designers' mindset to treat users of these systems as the weakest link.
This can be achieved by, for example, using security designs that are simpler and less error-prone, while considering the worst-case scenario that the users may have a high cognitive load when using these systems.
How to quantify users' vulnerability to social engineering attacks is an outstanding problem because it paves the way to quantify the cybersecurity gain of a better design when compared with a worse design.

\item Designing effective training to enhance users' self-efficacy. Training is an important mechanism for defending against social engineering cyberattacks. 
However, it is a challenging task to design effective training. 
One approach to addressing the problem is to routinely expose users to specific socially engineering cyberattacks (e.g., messages that have been used by attackers). Another approach is to insert sanitized social engineering attacks (e.g., phishing emails without malicious payload) into users' routine activities to trigger users' response to social engineering cyberattacks (e.g., the feedback will point to the user in question whether the user correctly processed the message).
This also effectively increases users' perception of social engineering cyberattacks, making them appear more frequent than it is.

\item Understanding and quantifying the impact of short-term factors in social engineering cyberattacks. We observe that as discussed above, few studies have examined the effect of short-term factors in social engineering cyberattacks. Short-term factors are known to affect cognition and behavior in other contexts profoundly.   
As highlighted in Eq. \eqref{eq:quantitative-model} and discussed above, we stress the importance of defining and quantifying social engineering cyberattack metrics, which are largely missing and will become an indispensable component of the broader family of cybersecurity metrics as discussed in \citet{pendleton2016acm,10.1145/3277666,XuBookChapterCD2019}.
\end{itemize}

Table \ref{table:future-research-directions} highlights the five future research directions and their relationships to the insights.

{
\begin{table}[!htbp]
\centering
\begin{tabular}{p{.3\textwidth}|p{.65\textwidth}}
 \hline
 \textbf{Future Research Direction} & \textbf{Insights}\\
 \hline\hline
 Human-Machine interactions &  Reduce trust (Insights \ref{insight:workload}, \ref{insight:stress}, \ref{insight:culture}, \ref{insight:message-quality-and-appeal}); Increase suspicion (Insights \ref{insight:vigilance}, \ref{insight:domain-knowledge}, \ref{insight:self-efficacy}, \ref{insight:age})   \\
  \hline
 Immunity to social engineering cyber attacks & Identify  and quantify underlying causes of immunity (Insights \ref{insight:domain-knowledge}, \ref{insight:self-efficacy}, \ref{insight:age}); Security and UI design (Insight \ref{insight:workload}, \ref{insight:stress}, \ref{insight:vigilance}); Improve message detection  (Insight \ref{insight:culture}, \ref{insight:persuasion-to-f}, \ref{insight:message-quality-and-appeal}) \\
  \hline
  Human-Centric System Design & Incorporate psychological state of computer user during system design (Insight \ref{insight:workload},  \ref{insight:stress}, \ref{insight:vigilance}, \ref{insight:attack-frequency}) \\ 
\hline
 Designing effective training  & Training based on susceptibility elements (Insight \ref{insight:personality}, \ref{insight:awareness}, \ref{insight:gender}, \ref{insight:culture}, \ref{insight:attack-frequency}, \ref{insight:training-to-rescue}, \ref{insight:message-quality-and-appeal}, \ref{insight:defense})\\
 \hline
 Understanding and quantifying the impact of short-term factors & Increase research focus on short term factors impact in security (Insight \ref{insight:workload}, \ref{insight:stress}, \ref{insight:vigilance})\\
 \hline
 \end{tabular}
\caption{Relationships between future research directions and insights.}\label{table:future-research-directions}
 \end{table}
}

\subsection{Further Discussion}
\label{subsection:discussion}

First, we observe that the 15 insights mentioned above are all {\em qualitative}, rather than {\em quantitative}. Moreover, the factors are typically investigated standalone.
Furthermore, even the qualitative effects are discussed in specific scenarios, meaning that they may not be universally true. Summarizing most of the insights mentioned above, the state-of-the understanding is: (i) cognitive workload, stress, and attack effort, increases one’s vulnerability to social engineering cyberattacks; (ii) the effect of vigilance, personality, awareness, culture remains to be investigated to be conclusive; (iii) domain knowledge, (certain kind) experience, age (together with certain other factors) reduces one’s vulnerability to social engineering cyberattacks; and (iv) gender may not have a significant effect on one’s vulnerability to social engineering cyberattacks. 

\begin{figure}[!htpb]
    \centering
    \includegraphics[width=.96\textwidth]{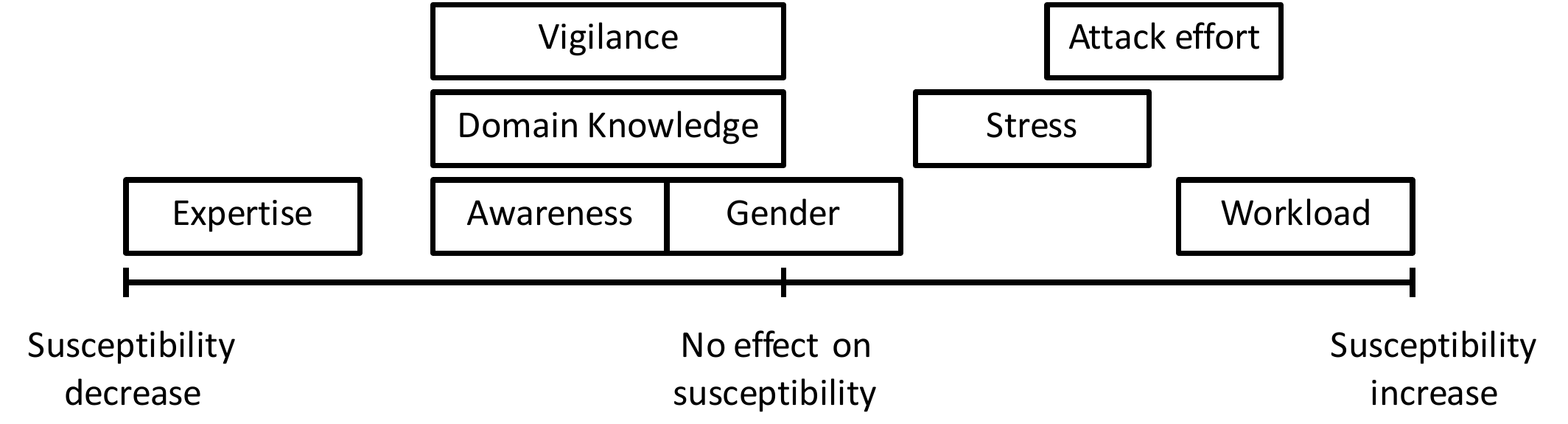}
    \caption{Our {\em speculation} of the impact on one’s susceptibility to social engineering cyberattacks: a factor on the left-hand side means that substantially increasing the factor (e.g., expertise) will  decrease one's susceptibility, with further left indicating a bigger degree in decreasing susceptibility; a factor on the right-hand side means that substantially increasing the factor (e.g., workload) will increase one's susceptibility, with further right indicating a bigger degree in increasing susceptibility; gender has little or no effect on one's susceptibility.}
    \label{fig:factor-importance-scale}
\end{figure}

Figure \ref{fig:factor-importance-scale} depicts our {\em speculation} of the impact on one's susceptibility to social engineering cyberattacks. Specifically, we suspect that expertise can decrease one’s susceptibility to the largest extent among the factors because expertise equips one the capability to detect the deceptive cues that are used by social engineering cyberattacks. We suspect that vigilance and domain knowledge have a significant, but smaller, impact on reducing one’s susceptibility because it is perhaps harder for an expert to fall into victim of social engineering cyberattacks. Since there is no evidence to show which one of these two factors would have a bigger impact than the other, we subjectively treat them as if they have the same impact.  We suspect awareness would have a significant impact on reducing one’s susceptibility, despite that the literature does not provide any evidence. According to Insight \ref{insight:gender} (which is drawn from a body of literature reviewed above), gender has little or no impact. We suspect that stress would decrease one’s capability in detecting deception cues, but attack effort would have an even more significant impact on increasing one’s susceptibility. We suspect that workload may have the biggest impact on one’s susceptibility because it would substantially reduce one’s ability in detecting deception cues. For other factors like age and culture, we suspect that their impacts might have to be considered together with other factors, explaining why we do not include them in Figure \ref{fig:factor-importance-scale}.
In summary, our understanding of the factors that have impacts on human's susceptibility to social engineering cyberattackers is superficial. This was indeed one of our motivations for proposing the mathematical framework outlined in Eq. \eqref{eq:quantitative-model}.

Second, it is a fascinating research problem to fulfill the quantitative framework envisioned in the paper because its fulfillment will permit us to identify cost-effective, if not optimal, defense strategies against social engineering cyberattacks. This will also help identify the most important factors. However, we suspect that the optimal defense strategies will vary with, for example, different combinations of short-term factors and long-term factors. For example, we suspect that the importance of factors may be specific to attack scenarios. This is supported by two very recent studies:  \citet{van2019arXiv} 
observed that certain short-term and long-term factors (e.g., workload) may be exploited to wage phishing attacks because malicious emails can coincide with high email volume;
and \citep{jalali2020jmir} showed that certain short-term and long-term factors (e.g., high workload and lack of expertise) are two important factors against medical workers. For example, Insight \ref{insight:awareness} says that awareness and general technical knowledge do not necessarily reduce one’s susceptibility to social engineering cyberattacks; however, this may not hold when taking awareness and human cognition functions into consideration. In other words, we can speculate that the effect of considering one factor alone and the effect of considering multiple interacting factors together may be different. This phenomenon is also manifested by Insight \ref{insight:self-efficacy}, showing that self-efficacy, knowledge, and previous encounter of social engineering cyberattacks collectively reduce one’s susceptibility to social engineering cyberattacks. In particular, costly phishing experiences would greatly reduce one’s susceptibility to social engineering cyberattacks, while non-costly experiences do not. Putting another way, a certain previous encounter may or may not have a big effect when considered together with other factors.

Third, Insight \ref{insight:training-to-rescue} says that effective training should not ask people to consciously think about social engineering cyberattacks, but making people to formulate an unconscious habit in coping with these attacks. This points out an important research direction on designing future training systems.

Fourth, last but not the least, studies in the context of social engineering cyberattacks inevitably involve human subject, meaning that ethical aspects of these studies must be taken into adequate consideration when designing such experiments and an IRB approval must be sought before conducting any such experiment. For ethical considerations in phishing experiments, we refer to \citep{4135777} for a thorough treatment.

\section{Conclusion} 
\label{s.concl}

We have presented a framework for systematizing human cognition through the lens of social engineering cyberattacks, which exploit weaknesses in human's cognition functions. The framework is extended from the standard cognitive psychology to accommodate components that emerge from the cybersecurity context. In particular, the framework leads to a representation of a victim's behavior, or more precisely, the degree a victim is persuaded by an attacker to act as the attacker intended, as some mathematical function(s) of many aspects, including victim's cognition functions and attacker's effort. We articulate a number of research directions for future research. We hope that this mathematical representation will guide future research endeavors towards a systematic and quantitative theory of Cybersecurity Cognitive Psychology.

\smallskip

\noindent{\bf Acknowledgement}. We thank the anonymous reviewers for their constructive comments, which have guided us in improving the paper. In particular, the entire Section \ref{subsection:discussion} is inspired by the reviewers' comments.

\section{Funding}
This paper was partially funded through a Mitre Basic Education Assistance Program (BEAP). The funders had no role in study design, data collection and analysis, decision to publish, or preparation of the manuscript.

\bibliographystyle{unsrtnat}
\bibliography{biblio}

\end{document}